\documentclass[12pt]{article}\usepackage[hyperfootnotes=false]{hyperref}
   \usepackage{epsfig}
      \usepackage{amsmath}
      \usepackage{amssymb}
  \usepackage{graphicx}
  \newlength{\abstractwidth}
  \renewcommand{\thefootnote}{\fnsymbol{footnote}}
  \renewcommand{\thanks}[1]{\footnote{#1}} 
  \newcommand{\starttext}{
  \setcounter{footnote}{0}
  \renewcommand{\thefootnote}{\arabic{footnote}}}
  \renewcommand{\theequation}{\thesection.\arabic{equation}}
  \newcommand{\be}{\begin{equation}}
  \newcommand{\bea}{\begin{eqnarray}}
  \newcommand{\eea}{\end{eqnarray}}
  \newcommand{\beq}{\begin{equation}}
  \newcommand{\ee}{\end{equation}}
  \newcommand{\eeq}{\end{equation}}

  \def\ba{\begin{eqnarray}}
  \def\ea{\end{eqnarray}}

  \def\12{{1 \over 2}}
  \def\eq{&=&}

  \def\d{\partial}

  \def\la{\langle}
  \def\ra{\rangle}

  \def\simleq{\; \raise0.3ex\hbox{$<$\kern-0.75em
      \raise-1.1ex\hbox{$\sim$}}\; }
   \def\simgeq{\; \raise0.3ex\hbox{$>$\kern-0.75em
      \raise-1.1ex\hbox{$\sim$}}\; }

\def\ba{\bf{a}}

  \def\h3{{\cal{H}}_3}

\def\o3{\Omega_3}

\def\O2{\Omega_2}
\def\o{\omega}
 \def\d2{$dS_{2+1}$}

 \def\bi{\begin{itemize}}
  \def\ei{\end{itemize}}

      \def\2{   $2$-adic}

\begin{document}
  \renewcommand{\theequation}{\thesection.\arabic{equation}}

\begin{titlepage}
  \rightline{}
  \bigskip

  \bigskip\bigskip\bigskip\bigskip

    \centerline{\Large \bf { }}
    \bigskip

    \centerline{\Large \bf { Singularities, Firewalls, and Complementarity }}
    \bigskip

  \bigskip \bigskip

  \bigskip\bigskip
  \bigskip\bigskip

  \begin{center}
  {{ Leonard Susskind}}
  \bigskip

\bigskip
Stanford Institute for Theoretical Physics and  Department of Physics, Stanford University\\
Stanford, CA 94305-4060, USA \\

\vspace{2cm}
  \end{center}

  \bigskip\bigskip

 \bigskip\bigskip
  \begin{abstract}


Almheiri, Marolf, Polchinski, and Sully, recently claimed that once a black hole has radiated more than half its initial entropy (the Page time), the horizon is replaced by a ``firewall" at which infalling observers burn up, in apparent violation of  the equivalence principle and the postulates of black hole complementarity. In this paper I review the arguments for firewalls, and give a slightly  different interpretation of them.  According to this interpretation the horizon has standard properties, but the singularity is non-standard.  The growing entanglement of the black hole with  Hawking radiation causes the singularity to  migrate toward the horizon, and eventually intersect it at the page time. The resulting collision of the singularity with the horizon leads to the firewall. Complementarity applies to the horizon and not to the singular firewall.

Almheiri, Marolf, Polchinski, and Sully   conjecture that firewalls form much earlier then the Page time; namely at the scrambling time.  I argue that there is no reason to believe this generalization, and good reason to think it is wrong.

For most of this paper I will assume that the firewall argument is correct. In the last section before the conclusion I will describe reasons for having reservations.

 \medskip
  \noindent
  \end{abstract}

  \end{titlepage}

  \starttext \baselineskip=17.63pt \setcounter{footnote}{0}

\tableofcontents

\setcounter{equation}{0}
\section{Complementarity}

If correct, the firewall phenomenon \cite
{Almheiri:2012rt} represents a serious departure from our previous understanding of black holes, and possibly other kinds of horizons or apparent horizons. Potentially it could undo a great deal of thinking about de Sitter space, inflation, and eternal inflation.

This is especially true if, as argued in \cite{Almheiri:2012rt}, the time scale for forming firewalls is the scrambling time and not the Page time ( the time at which half the entropy of the black hole has evaporated away). Thus it is important to get it right.

Just to give a sense of the orders of magnitude, the Page time for a solar mass black hole is $10^{72}$ seconds, whereas the scrambling time is $10^{-1}$ seconds. For a black hole with Schwarzschild radius equal to the radius of a proton, the two time scales are the age of the universe and $10^{-20}$ seconds.

The firewall argument is not  conclusive and I can't tell if it's correct. My purpose is more modest: be  generous about the assumptions,  and then explore the conceptual implications of firewalls.

The following postulates are usually described as black hole complementarity. \cite{Susskind:1993if} \cite{Susskind:1993mu}.

 \bi

 \item
 Postulate 1: The process of formation and evaporation of a black hole, as viewed by
a distant observer, can be described entirely within the context of standard quantum theory.
In particular, there exists a unitary S-matrix which describes the evolution from infalling
matter to outgoing Hawking-like radiation.

\item

 Postulate 2: Outside the stretched horizon of a massive black hole, physics can be
described to good approximation by a set of semi-classical field equations.

\item  Postulate 3: To a distant observer, a black hole appears to be a quantum system with
discrete energy levels. The dimension of the subspace of states describing a black hole of
mass M is the exponential of the Bekenstein entropy $S(M)$.

\item
 Postulate 4:
A freely falling observer experiences
nothing out of the ordinary when crossing the horizon until the singularity is approached. Another way to say this is that no observer ever detects a violation of the known laws of physics. An example that has been studied is the apparent paradox of information cloning behind a horizon.

\ei

In \cite{Almheiri:2012rt}  Almheiri, Marlof, Polchinski, and Sully argue that these postulates are not mutually consistent, at least for old enough black holes \footnote{Long ago N. Itzhaki made related claims although in that case the argument did not refer to the age of the black hole \cite{Itzhaki:1996jt}.}. The argument leads them to postulate firewalls at which infalling observers burn up before they reach the horizon. There has been confusion about the following question:

\bi
\item Does a firewall mean that infalling observers experience something abnormal on their way in, or  are they suddenly ``terminated"  without warning at the horizon.
\ei

 I will begin with that issue.

\setcounter{equation}{0}
\section{Firewalls}
\subsection{What the Argument Says}

Let's consider an experiment involving three systems $A,$ $B,$ and $R,$ each composed of some large number $N$ of distinguishable spins. There are two possible ways that the spins have been prepared. In the first way, $A$ and $B$ are maximally entangled. For example each spin in $A$ may  be entangled with a partner in $B$ in a singlet state.  The system $R$ is in its own pure state which is unentangled with either of the other two systems. Let's call this state of $A, \ B,$ and $R,$ the ``safe" state.

The other possibility is that $B$ is maximally entangled with $R,$ and $A$ is in a pure state.
Call this state the ``dangerous" state. Note that general quantum mechanical principles insure that $B$ cannot be maximally entangled with both $A$ and $R$. If fact the greater the degree of entanglement of $(B,R)$ the less the entanglement of $(B,A).$

Alice is going to do an experiment in which she first makes a series of measurements on $B,$ followed by  a series of measurements on $A.$ For example, she may measure the $z$ components of the spins in $B,$ and then do the same in $A.$ The rule is that if she discovers evidence that $A$ and $B$ are NOT entangled (dangerous state), she blows herself up. For example she might discover that some of the pairs have parallel spins whereas in the safe state they must be anti-parallel.

If, on the other hand, the evidence is consistent with the safe $(A,B)$ entangled state, she does nothing out of the ordinary. If $N$ is large enough she can make these determinations with high fidelity.

On the other hand, suppose that before Alice makes any measurement on $B,$ she decides to test the situation by making measurements on $R.$ Once she has measured both $B$ and $R$ she knows, with high fidelity, whether the state is safe or dangerous. If it's dangerous she can just decide to stop the  experiment.
In other words after measuring $B,$ Alice's state of knowledge will be different depending on whether she has or has  not previously measured $R.$

The analog to the question posed in the previous section is:

\bi

\item Between the time Alice measures $B$ and $A,$ is there any observable difference between the two situations; safe and dangerous?
\ei

The answer is obvious in this case. If she does not measure $R,$ her state after measuring $B$ is exactly the same for the two cases. But if she does measure $R$ she easily can tell the difference.

Now consider Alice falling into a black hole. She does not know when the black hole formed or how much Hawking radiation has been emitted. In particular she does not know if the black hole is younger or older than the Page time (young or old).

 Let us suppose that the black hole is in fact old. According to the arguments of \cite{Page:1993wv}, and more recently Hayden and Preskill  \cite{Hayden:2007cs}, by that time the black hole will be maximally entangled with the outgoing radiation. This means that any subsystem of the black hole will be maximally entangled with the outgoing quanta. I will accept that as correct.

\begin{figure}[h!]
\begin{center}
\includegraphics[scale=.3]{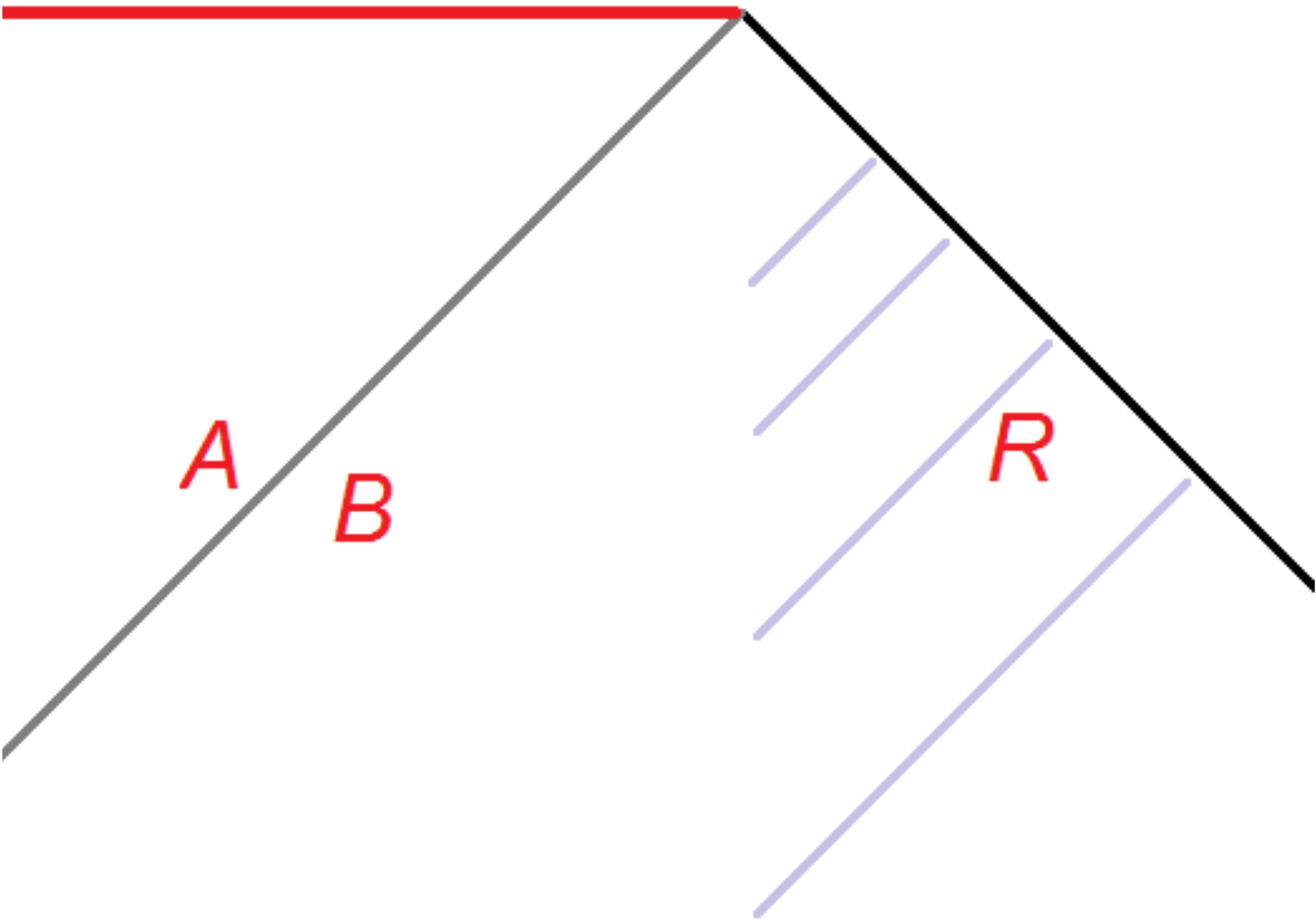}
\caption{A small portion of an old black hole. $A$ and $B$ are points on opposite sides of the horizon that the infalling observer sees as neighboring. R is the outgoing Hawking radiation.}
\label{f1}
\end{center}
\end{figure}

The symbols  $A, \ B,$ and $R$ will now stand for the following:

$B$ represents the degrees of freedom in a radiation field just outside the horizon, far enough away that low energy quantum field theory applies (a region we will later call the zone). $A$ represents modes just inside the horizon.  $R$ represents the Hawking radiation that has been emitted up until the time of the experiment. If the black hole is young then $A$ and $B$ are very entangled, in about the same way that they would be in flat space. $B$ would not be entangled with $R$.

Let's suppose that $A$ and $B$ are unentangled, i.e., very different from the Minkowski vacuum. Then conditions in the region between them must be highly excited and therefore dangerous. So Alice's safety, when passing the horizon, depends on strong entanglement between $A$ and $B.$ In other words she is safe if the black hole is young.

If the black hole is old then by Page's argument \cite{Page:1993wv} \cite{Hayden:2007cs} $B$ is maximally entangled with $R,$ and not with $A.$ In this case Alice will be terminated by a firewall.

As before, if she does not make a measurement on $R,$  Alice's experiences, right up to the point where she crosses the stretched horizon are the same for either case, young or old. In particular she does not encounter any unusual events as she crosses $B,$  and interacts with it.

But Alice has access to the Hawking radiation $R$ before she jumps in. If $R$ is maximally entangled with $B,$ Alice can detect that fact by studying correlations between the two. Thus,  after crossing $B,$ she will know if the state is dangerous or safe.

 The the answer to the question is again obvious: If Alice does not measure $R$ there will be absolutely no detectable difference(between the old and young black holes) between the time she passes $B$ and the time she gets to the horizon.

One might argue that all this is based on a red herring; that it is mixing two complementary descriptions which is causing confusion. Indeed $B$ and $R$ are on the outside of the horizon and $A$ is on the inside. Do we have any right to consider them in the same quantum description?

 In fact the argument can be made in Alice's infalling frame where all three systems are in her past causal patch. Just because someone on the outside cannot see in, that does not mean that someone on the inside cannot see out. If the black hole is old enough, then enough radiation quanta are in Alice's past so that the argument can be made in her frame.

 There is one possible obstruction to Alice carrying out the experiment to test $R$ before she falls in. The type of experiment she has to do is exceedingly complicated and involves simultaneously manipulating almost all the radiation quanta. It is possible that this cannot be done by any experiment before the black hole has evaporated\footnote{This possibility has also been noticed by Bousso. Private communication.}. This may be the weakest link in the argument against complementarity. I will ignore this issue in what follows.

\subsection{Low and High Angular Momentum}

The black hole geometry contains a near-horizon-region: following Bousso I will call it the \it zone. \rm The zone is the
 portion  of space between the stretched horizon and a distance of order $R_s$ from the horizon. The zone is also the region
 that can be approximated by Rindler space.  The proper temperature varies in the zone from near Planckian, to the Hawking temperature. As long as we keep away from the Planckian end, this region should be describable by ordinary quantum field theory. The entropy  stored in this portion of space is part of the Bekenstein entropy, and although it is a small fraction of the total, it contains enough heat to be dangerous to anyone suspended above the horizon.
Furthermore it is proportional to the area of the horizon.

This zone-entropy is distributed over all angular momenta from $l=0$ to   $l=R_s m_p,$ where $R_s$ is the Schwarzschild radius and $m_p$ is the Planck mass. The higher the angular momentum, the closer the modes are to the horizon. The correct picture is that the high $l$ quanta are emitted and absorbed by the stretched horizon and thereby thermalized.

On the other hand the modes which appear as real Hawking radiation are mainly low angular momentum. The occupation numbers of higher modes are exponentially suppressed by tunneling barriers that separate the zone from the outside. For simplicity I will refer to the real Hawking radiation as S-wave radiation.

The most rigorous argument in \cite{Almheiri:2012rt} uses Postulate 1 in an essential way.
 The authors refer to it as the "purity of Hawking radiation."
Since  the actual outgoing quanta in Hawking radiation are primarily low angular momentum quanta, the S-matrix argument applies to these modes and not directly to the vast reservoir of high angular momenta degrees of freedom that comprise most of the entropy of the black hole. On the other hand, the low angular momentum degrees of freedom are very dilute. The black hole emits one S-wave quantum every Schwarzschild time, and that quantum is spread over the entire horizon area. Even if the S-wave degrees of freedom are completely entangled with the radiation, Alice would probably not be seriously affected by them. To make the argument that there is a dangerous firewall the degrees of freedom of $B$ must include the high angular momentum modes near the horizon. It is difficult to see how the S-matrix by itself can access these modes. Therefore the purist S-matrix viewpoint is not sufficient to demonstrate that old black holes have dangerous firewalls.

There are two alternative arguments. Reference \cite{Almheiri:2012rt} argues that one can replace Hawking evaporation by a man-made mining operation on the black hole. One sends down buckets and collects the high angular momentum quanta, then pulls them out. There may be many holes in this bucket, and in any case it does not say what happens if you don't mine the black hole.

The other  argument is mathematical. It says that all the radiation modes in the zone are thermalized at Rindler temperature $\frac{1}{2\pi}.$ All the modes have the approximately same Rindler frequency; in dimensionless Rindler units the frequencies are order $1$ and the energy is equipartitioned. From that it  follows that all modes are maximally entangled with the Hawking radiation $R.$ Thus $B$ can correspond to a high angular momentum degree of freedom.

If the argument can be made rigorously for the S waves, it would seem strange that it would not apply also to the high angular momenta.

\section{Firewalls are Part of the Singularity}

\subsection{The Non-existence of Space Behind a Firewall}

\begin{figure}[h!]
\begin{center}
\includegraphics[scale=.3]{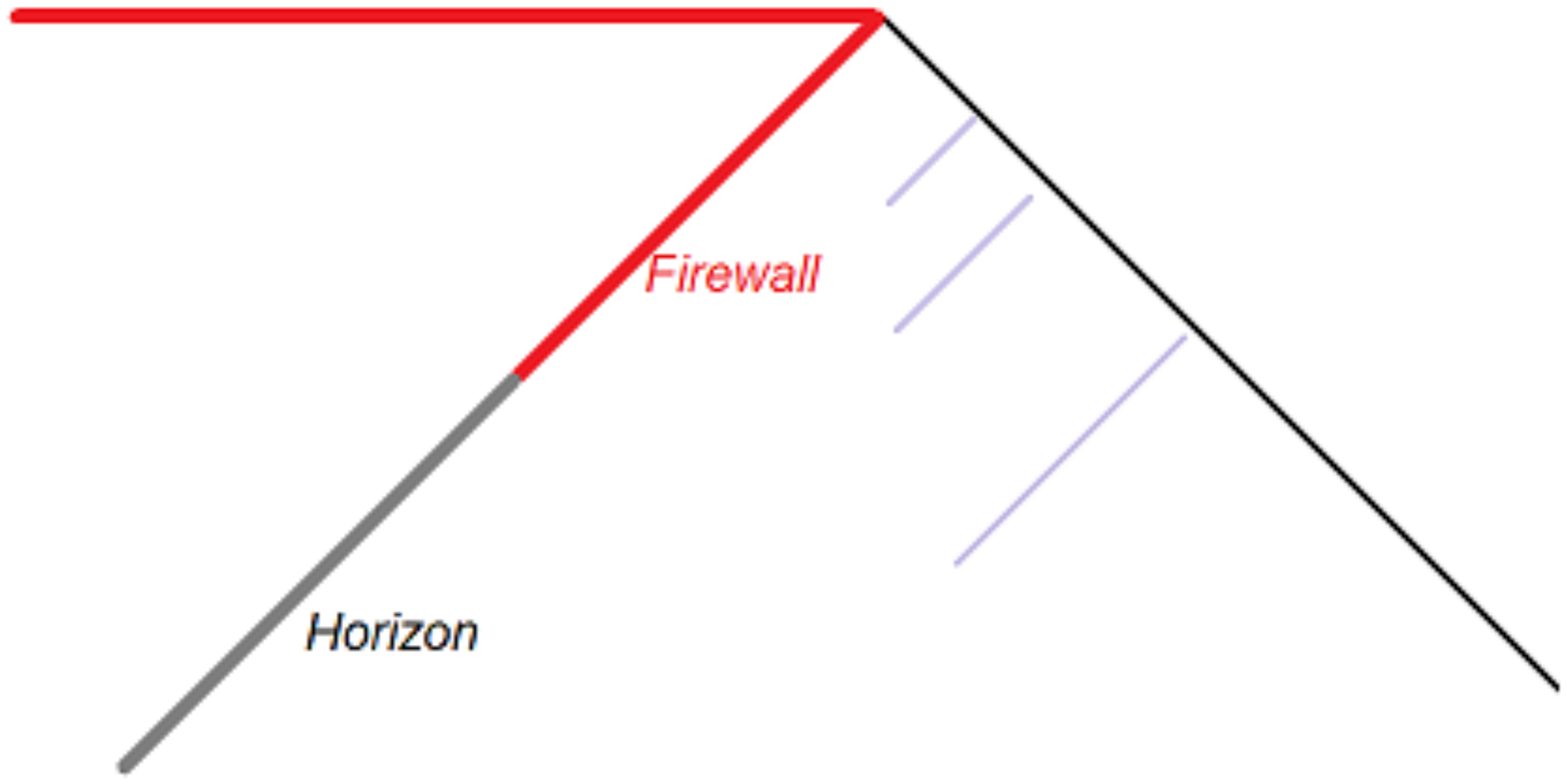}
\caption{The formation of a firewall at the Page time.}
\label{f2}
\end{center}
\end{figure}

Let's consider how the  Penrose diagram  for an old black hole might be modified due the firewall phenomenon. Figure \ref{f2}
shows a possible configuration in which the firewall originates on the horizon at the Page time. Before that the horizon is conventional.

Firewalls, if they exist, are disruptive of the smooth Rindler-like  horizon of a black hole.  In fact the argument can be made that the lack of entanglement between the inside and the outside modes demonstrates something stronger.
Czech,  Karczmarek,  Nogueira, and Van Raamsdonk
\cite{Czech:2012be} \cite{Czech:2012bh} (see also the review \cite{Mathur:2012zp}), argue that the lack of entanglement of the two sides of the horizon means the space behind the horizon does not exist at all. The conclusion of this type of argument is not only that the horizon is polluted, but the entire space-time behind a firewall does not exist.
The problem with
Figure \ref{f2} is that it is not consistent with the idea that there is no space behind the firewall. An observer can cross the conventional horizon and migrate to the region behind the firewall. Moreover that observer would see the origin of the firewall as a naked singularity.

\begin{figure}[h!]
\begin{center}
\includegraphics[scale=.3]{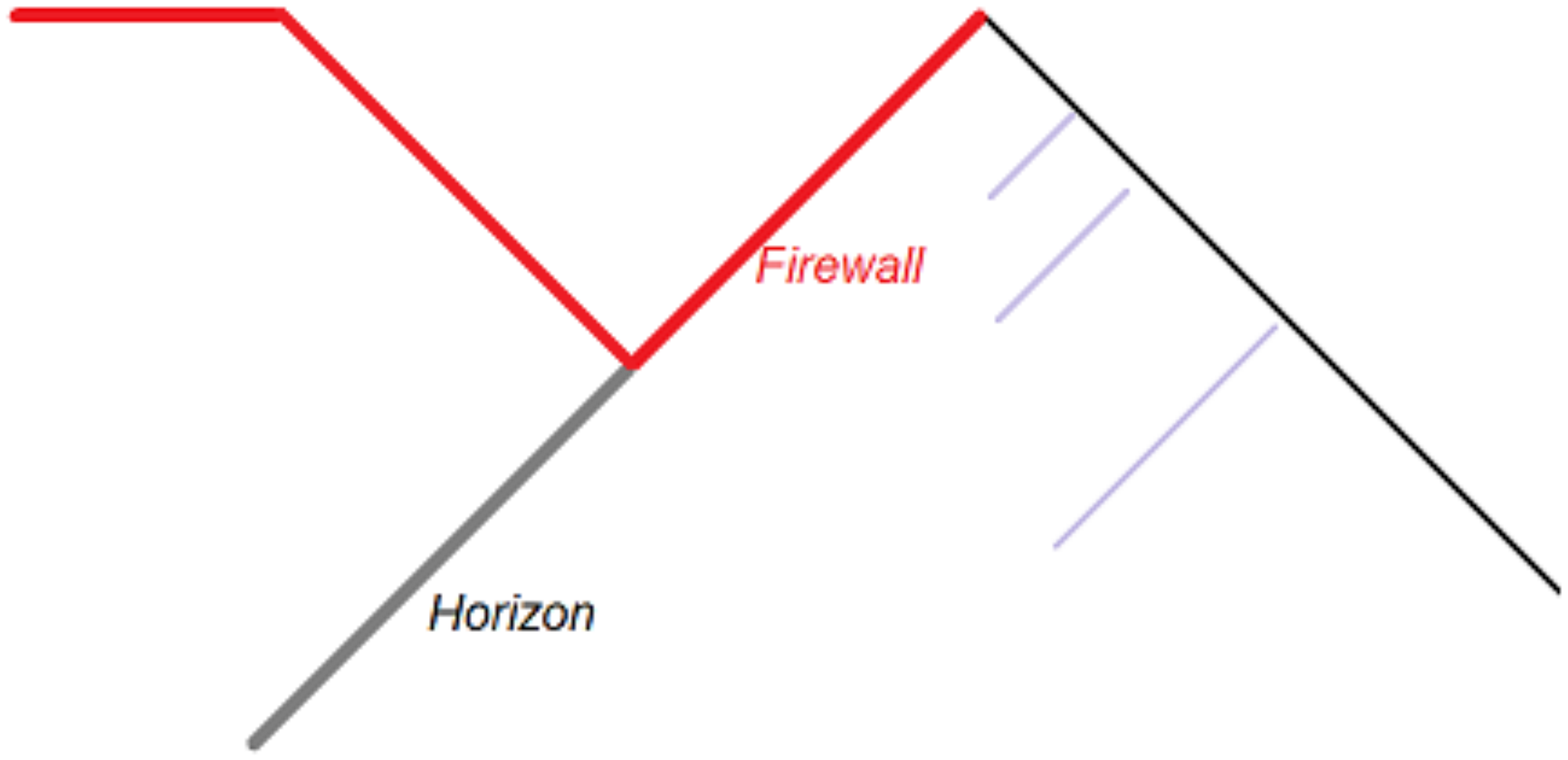}
\caption{The region behind the firewall does not exist. The firewall is an extension of the singularity.}
\label{f3}
\end{center}
\end{figure}

A diagram which is more consistent with the hypothesis that the firewall is the end of  spacetime is shown in Figure \ref{f3},

\subsection{A More Gradual Picture}

\begin{figure}[h!]
\begin{center}
\includegraphics[scale=.3]{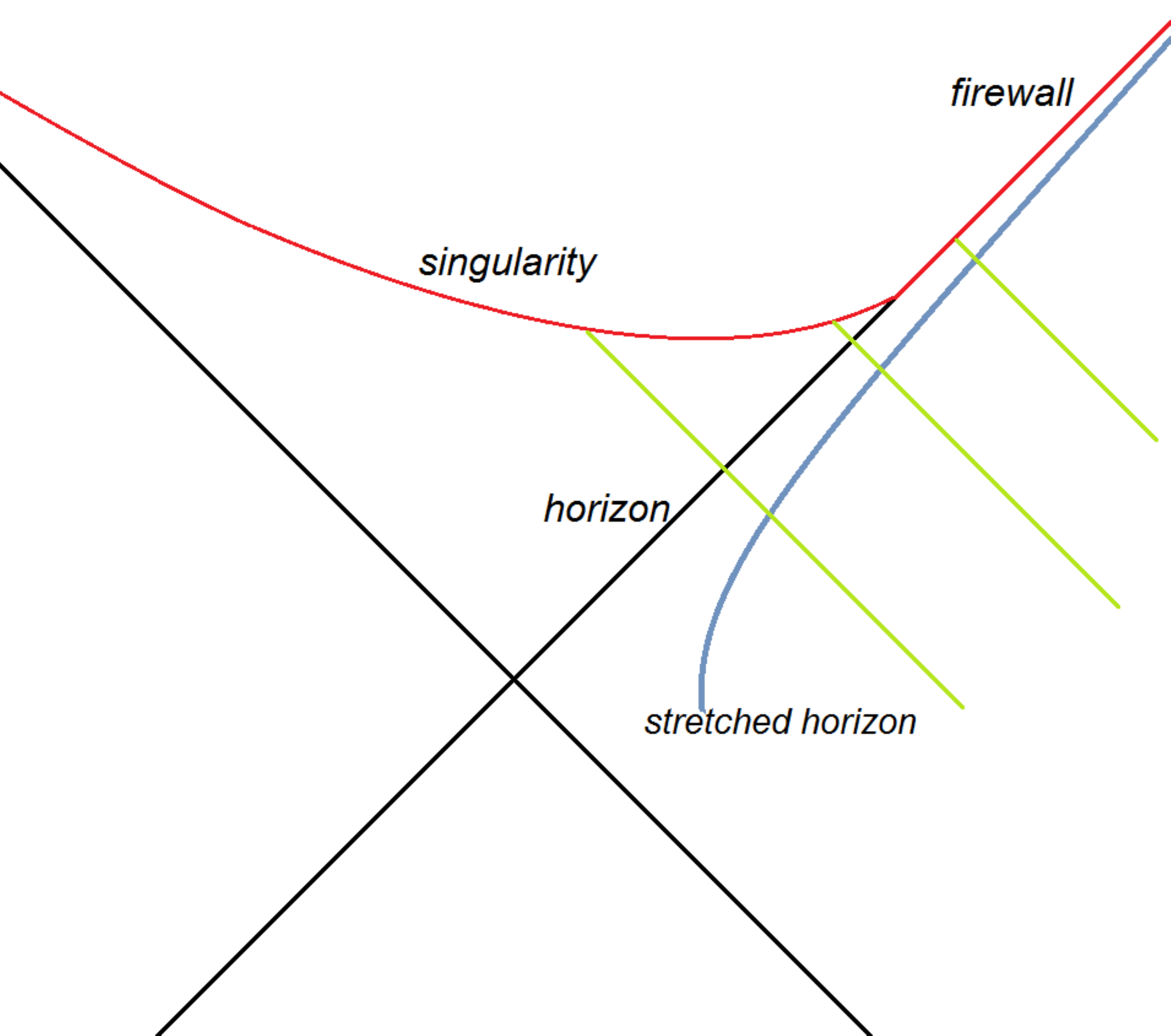}
\caption{The same as Figure \ref{f3} but with the singularity smoothed and spacelike.}
\label{f4}
\end{center}
\end{figure}

The suddenness of the onset of the firewall in Figure \ref{f3} is unlikely. The  high degree of entanglement between $B$ and $R$ does not happen suddenly. A possible way to reflect the gradual onset before the Page time is  to smooth the trajectory of singularity and make it spacelike \footnote{I thank Eva Silverstein and Simeon Hellerman for discussions of this point.}. In Figure \ref{f4} this is shown in Kruskal coordinates.

Instead of thinking of the firewall as part of the horizon, Figure \ref{f4} suggests that we think of it as an extension of the singularity. The horizon consists of only the black portion of the light-sheet; the portion before the red firewall forms. With that interpretation, there is no conflict between postulates 2 and 4. Horizons, by definition, have no firewall.

Figure \ref{f4} shows a series of lightlike infalling trajectories that cross the stretched horizon, and intersect the singularity. Such trajectories can be used to define the location of the singularity as a function of time. With the smooth spacelike evolution, the separation of the singularity from the horizon (or stretched horizon) is a continuous function of time that goes to zero at the Page time.

If this picture is correct it must be possible to construct a quantitative theory of the evolution of the singularity.

\subsection{Conservation of Entanglement}

Coarse-grained entropy is defined in such way that it is additive. The total coarse-grained entropy of a black hole plus the Hawking radiation at any time is given by,\
\be
S_{cg} = S_{bh} + S_r
\ee
where $S_{bh}$ is the Hawking Beckenstein entropy of the black hole. The entropy of the radiation is $S_r.$ It is essentially the number of quanta that have been emitted. The second law of thermodynamics says that $S_{cg}$ does not decrease. For the evaporation of a black hole it is approximately constant.
\be
 S_{bh}(t) + S_r(t)  = S_0
 \label{sss}
\ee
where $S_0$ is the  entropy of the initial black hole.

Unless Alice measures $R,$ her experiences interacting with $B$ do not distinguish the safe and dangerous states.  In the black hole context $B$ is the zone and her experiences in this region are the same for old and not-so-old black holes. Entanglement with $A$ and entanglement with $R$ lead to the same density matrix for $B.$

Our discussion up to now suggests the following modification of the spin model in Section 2.1.
Initially there are only two systems, $B$ and $A$ with maximal entanglement between them. Alice cannot access $A$ until she samples $B.$ There is nothing in $R.$
As time evolves, spins from $A$ are transferred to $R$ where Alice has access to them. The density matrix of $B$ is unchanged but the entanglement of $B$ is transferred from $A$ to $R.$ Eventually there is nothing left in $A$ and all the entanglement is with $R.$ One may say that there is a \it conservation of entanglement".\rm

In this picture, instead of blowing up, Alice finds fewer and fewer degrees of freedom after she passes B.  The argument of \cite{Czech:2012bh} would then say that there is no space behind the horizon for Alice to exist in.

\begin{figure}[h!]
\begin{center}
\includegraphics[scale=.3]{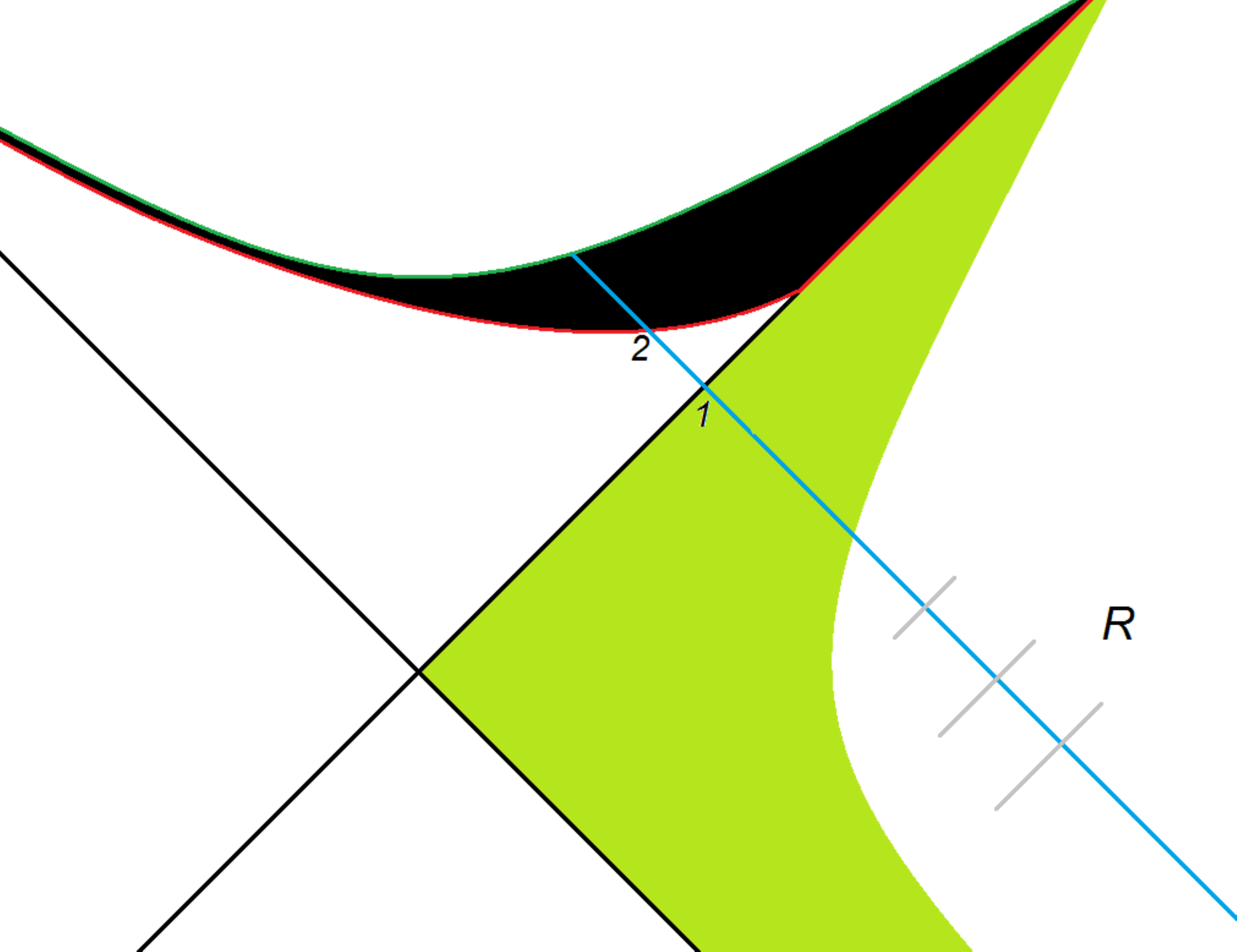}
\caption{The migration of the singularity. The black region is the part of the spacetime
that is removed by the shift of the singularity from the green curve to the red curve. The blue line is a light sheet the intercepts the Hawking radiation, the horizon, and the singularity.}
\label{f5}
\end{center}
\end{figure}

This model fits well with the evolution of the singularity. In Figure \ref{f5} the upper boundary of the black region is the location of the classical singularity and the lower (red) edge is the modified singularity which eventually becomes the firewall. The black region is excised from the geometry. A simple rule is that the the singularity occurs at the point (labeled $2$) such that
\be
[ \frac{A_1}{4} - \frac{A_2}{4} ]+ S_R = S_0
\ee
where $[ \frac{A_1}{4} - \frac{A_2}{4} ]$ is the covariant entropy bound on the blue light sheet
lying between the 2-spheres $1$ and $2,$ $S_R$ is the entropy in Hawking radiation passing the light sheet, and $S_0$ is the initial entropy of the black hole. The actual details are undoubtedly more complicated.

\setcounter{equation}{0}
\section{True and Apparent Horizons}

\subsection{Cloning}

It would seem that the problem of cloning has a much simpler and less subtle solution than described in \cite{Susskind:1993mu} and \cite{Hayden:2007cs}. Recall the setup: Alice crosses the horizon carrying a qubit: Bob hovers above the horizon until such time as he can recover Alice's bit from the Hawking Radiation: Then he jumps in: Meanwhile Alice sends her bit to Bob so that Bob sees the forbidden cloning of information. This would obviously violate Postulate 4. There are two cases that have been described.

\bi
\item Alice carries her bit in long before the Page time \cite{Susskind:1993mu}. In that case it is expected that Bob has to wait a retention time of order $M^3$ before jumping in. In order for Alice to send her message she must send it in the form of a quantum of exponentially large energy, $E=\exp{M^2}.$ If Alice possessed that much energy she would have drastically changed the location of the horizon.

    \item Alice jumps in after the Page time. In that case Hayden and Preskill \cite{Hayden:2007cs} argue that the retention time is the fast scrambling time $M \log{M}.$ Bob still jumps in too late to get Alice's message unless the message is sent with energy comparable to $M$. Again Alice would have drastically changed the location of the horizon. One finds in both cases the cloning cannot be witnessed by Bob. In the latter case, the protection against cloning is just enough, while in the former case it is vastly overkill.

\ei

With the new observation of firewalls the solution to the cloning problem seems much simpler, at least at first sight. In the case in which Alice goes in early, Bob has to wait at least the Page time to retrieve Alice's bit. By that time if he jumps in he is destroyed at the firewall.

In the second case Alice and her bit never get past the firewall.

I believe that this simple firewall explanation is not quite correct.

\subsection{Firewalls at the Apparent Horizon}

Let us consider  the second case in which Alice jumps after the Page time. The other case is similar.

\begin{figure}[h!]
\begin{center}
\includegraphics[scale=.3]{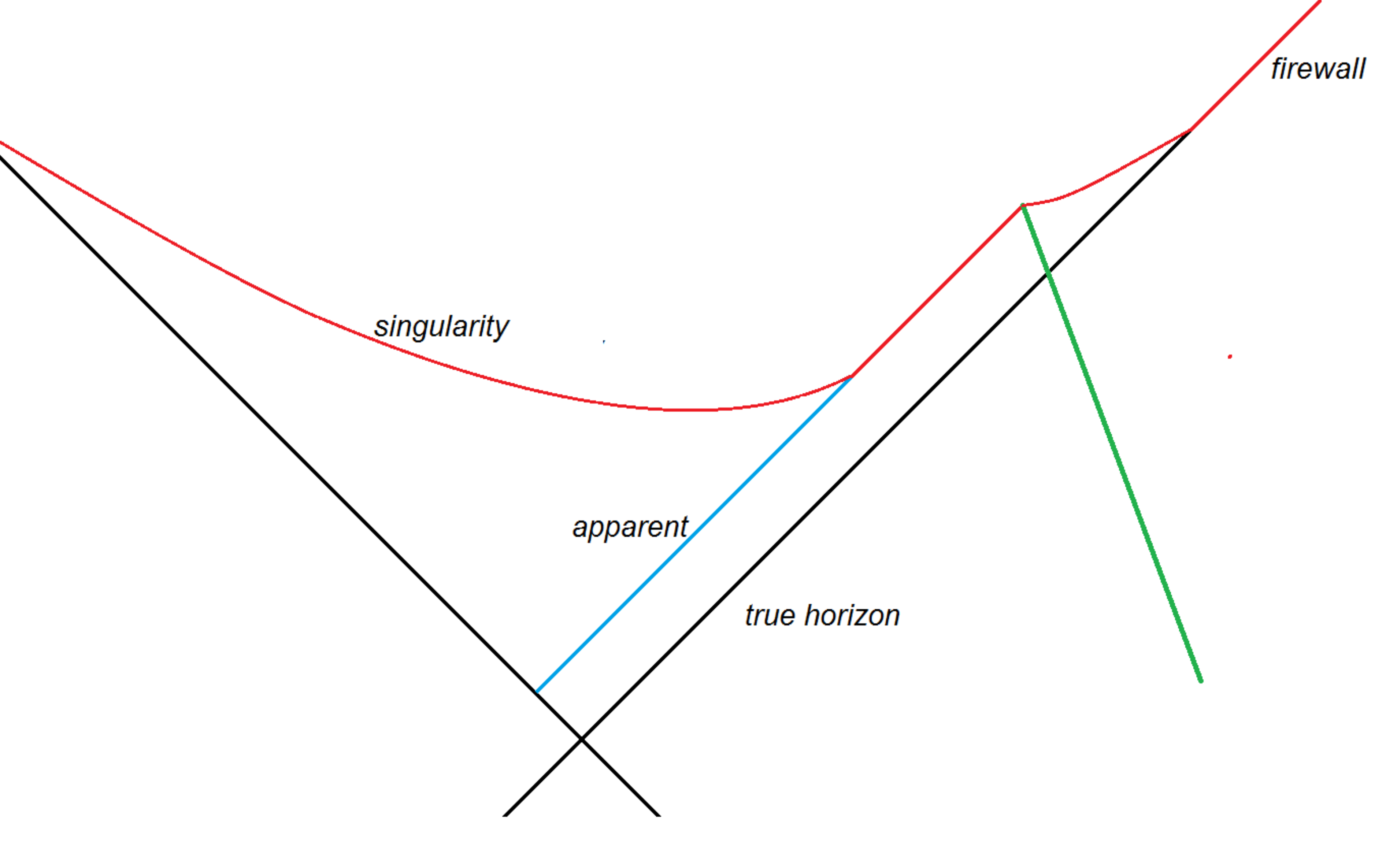}
\caption{When Alice jumps in, the true horizon shifts. The original horizon shown in blue is an apparent horizon. }
\label{f6}
\end{center}
\end{figure}

We will assume that Alice drops into the black hole from an orbiting space station at a drop-off distance $R_{DO}$ from the horizon. Alice has a mass $m_a$ which is much greater than the Planck mass. The result of Alice's fall is a shift outward of the true horizon. The original horizon is merely an apparent horizon, as in Figure \ref{f6}.
 One would describe the original  firewall as an extension of both the singularity and the apparent horizon, but not the true horizon.

Adding Alice's information to the black hole means that at first the black hole is no longer maximally entangled with the Hawking radiation. Therefore the true horizon is smooth when Alice jumps in. However, additional radiation together with the scrambling of Alice's bits, will eventually  maximally  entangle the black hole with the radiation. At that point a new firewall must form (also shown in Figure \ref{f5}). The time scale for forming the new firewall should be no less than the scrambling time for Alice's bits. However the scrambling time is not enough. During scrambling Alice's bits become entangled with the black hole, but not the radiation. That requires an additional time of order the number of bits that Alice brought in. Basically half those bits have to be radiated.

The conclusion is that the firewall did not prevent Alice's bit from arriving behind the true horizon. Therefore there is still a risk that Bob---assuming $ he $ can pass the horizon---may later see cloning.

\begin{figure}[h!]
\begin{center}
\includegraphics[scale=.3]{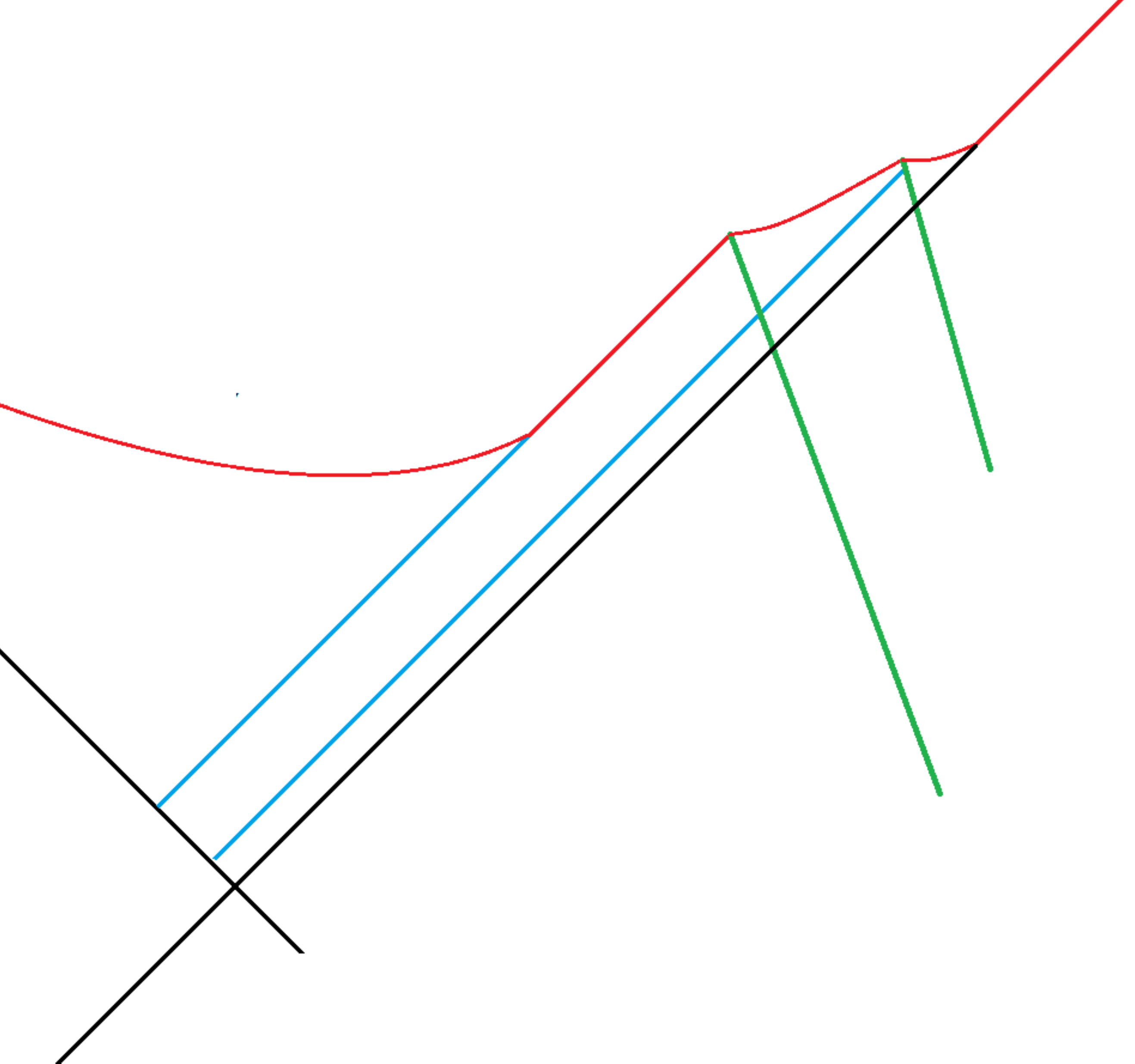}
\caption{Bob jumps in.}
\label{f7}
\end{center}
\end{figure}

 Clearly  Bob can also pass the horizon by the same argument. Bob's mass $m_B$ will also shift the horizon as in Figure \ref{f7}.

Evidently the firewall hypothesis does not in itself solve the cloning problem. However the arguments of \cite{Susskind:1993if} and \cite{Hayden:2007cs} still apply. Given that the singularity is much closer to the horizon than the classical expectation, it seems obvious that the firewall phenomenon makes the observation of cloning more difficult than estimated in \cite{Hayden:2007cs}.

\setcounter{equation}{0}
\section{Survival Times}

None of the above should be interpreted as saying that the firewall phenomenon, if it is real,  is less than remarkable. It is an extraordinary unexpected quantum effect that totally changes the nature of the singularity, shifting it from its usual classical  location to a location much closer to the horizon. If the analysis of \cite{Almheiri:2012rt} is correct, jumping into an old black hole is not safe, no matter how large it is. The actual horizon crossing may be uneventful but the infalling time until the singularity is extremely short.

To estimate the time that Alice has, we need to compute how long it takes for her world line to reach the apparent horizon. This is a classical calculation in which the Planck mass does not appear. Dimensional analysis would say that Alice's time is a function only of the black hole mass, Alice's mass, the distance from which Alice was dropped off, and the Newton constant \footnote{Don Marolf has pointed out that there can also be dependence on Alice's size. For simplicity I will assume Alice is a small as possible, i.e., Alice is a black hole.}. Calling Alice's survival time $T_A$
\be
T_A \sim m_a G F(R_{DO}, M/m_a)
\ee
where $F(R_{DO}, M/m_a)$ is a function of the drop-off distance, and the ratio of the black hole mass and Alice's mass.

Consider the case in which the drop-off distance is approximately Alice's Schwarzschild radius and take the Rindler limit $M\to \infty.$
 The result for $T_A$ should tend to a finite limit. Therefore for $M \gg m_a$ and $R_{DO} = m_a G$
\be
T_A \sim m_a G
\ee
For $m_a = 100 \ kg$ Alice would have about $10^8$ Planck times. That's small but it's not zero.

Dropping Alice from a distance equal to her own Schwarzschild radius is not realistic. Dropping her from a larger distance would mean that her momentum as she approached the horizon would be larger and therefore the perturbation of the horizon would be larger. The further away she is dropped from, the larger is her survival time $T_A.$

Alice could choose to jump in alongside a very large mass, and thereby increase $T_A.$ This  does not mean that she improves her total longevity but she does increase the time that she survives after  crossing the true horizon.

One thing to note is that the survival time is very sensitive to the mass of the infalling system. Ordinarily the opposite is true; the survival time is the classical geodesic distance from the point where the system crosses the horizon to the singularity. Typically this is of order $M,$  the mass of the black hole. However, since the horizon does respond to the infalling energy there is always some small dependence of that geodesic distance on the infalling mass.

In the case where the singularity includes the firewall, the mass of the black hole becomes  irrelevant.

To summarize, the firewall phenomenon does not prevent information from crossing the event horizon in the infalling frame. If, in the exterior frame, the information is in the Hawking radiation, then complementarity has to be invoked even after the Page time.

\setcounter{equation}{0}
\section{Page Time or Scrambling Time?}

The authors of \cite{Almheiri:2012rt} escalate the stakes radically by proposing that the time for formation of the firewall is not the Page time, but rather the Scrambling time. 

The argument that the scrambling time is the time-scale for firewalls is as follow. If the black hole is maximally entangled then its density matrix is maximally random. If something is true with a high degree of certainty for such a ensemble then it must be true for the overwhelming majority of pure states. Therefore firewalls must be present for almost all black hole states.  The argument then proceeds by claiming that the time for the state of a system to become typical is the scrambling time: that which is generic will be overwhelmingly likely after the scrambling time. Since firewalls are generic, they must be a feature of black holes after the scrambling time.

Before discussing the argument, we should note that it is not consistent with the conservation of entanglement discussed in Section 3.3. At the scrambling time the radiation entropy $S_R$ is negligible and $\frac{A_1}{4} \approx  S_0.$  Therefore $A_2$ must almost vanish and the singularity  is at its classical location

The fault in the argument is that the meaning of $scrambled $ is not at all the same as $ generic. $ For many properties of a system they are completely different. The reason for conflating the two is that for  most of the usual observables that are  experimentally accessible,  generic and scrambled $are$ the same. But there are reasons to think that these quantities do not include firewalls. Indeed there is something peculiar about observing a firewall. It cannot be done without passing through the horizon. From the point of view of the degrees of freedom, that describe a black hole in the outside frame of reference---matrix theory, superconformal field theory, brane constructions, fuzzballs---firewalls are evidently very difficult to describe by conventional observables. The reason, most likely, is that they are exactly the kind of observables for which scrambled and generic states are different concepts. We will begin by discussing various meanings of a generic state.


\subsection{Some Meanings of Generic}

Generic refers to what is true for the vast majority of states of a system. What is generic is what is true for a density matrix which maximizes the entropy, subject to whatever constraints may be relevant.

One definition of a generic state is that it is described by a completely incoherent density matrix proportional to the identity. An impure state of this type has equal probability for all basis states, no matter what basis is chosen. It is the density matrix with the absolute maximum entropy.
Of course such a state is never achieved since it corresponds to infinite temperature. But we can consider the density matrix with the maximum entropy for a given total energy. The corresponding thermal density matrix effectively truncates the space of states when the individual degrees of freedom have energy greater than the temperature. But within the truncated space a thermal density matrix is close to a completely incoherent state.

Any pure state is non-generic for some quantities. To illustrate, let us consider a large macroscopic system in a pure state.
Let the energy levels of the system be labeled $n$ with energy $E_n,$ and let the corresponding eigenvectors be $|n\rangle.$ A general pure state has the form
\be
\Psi(0) = \sum_n F_n |n\rangle.
\label{11}
\ee
Now consider a small part of the system and trace out over the rest. The small subsystem will be described by a thermal density matrix with a temperature which is chosen to reproduce the average energy in the small subsystem. In other words the state of the small subsystem is maximally incoherent subject to the constraint. According to arguments pioneered by Page, a small subsystem can mean anything up to half the number of degrees of freedom of the total system. The entropy of the small (meaning less than half) subsystem is maximal until the size of the system exceeds half the total system. Once the subsystem exceeds the half way point, the entropy (in the case of the overall pure state) starts to decrease. This phenomenon of maximal entropy for all small subsystems is called scrambling. In a scrambled state almost everything that we normally measure has the generic thermal value. That is because the things we measure usually can be constructed from the observables of  small subsystems.

On the other hand there are global observables  which generally do not exhibit generic behavior. These are not the usual things we measure and they depend on the details of the pure state \ref{11}. Whether they are generic or not cannot be determined on the basis of whether the system is scrambled, for the simple reason that the definition of scrambling only involves small subsystems. Typically they involve at least half the degrees of freedom in an extremely intricate way. These global observable do not become generic in a scrambling time.

In analyzing  the time scales for firewall formation we may or may not have to take into account the evaporation process. If we want to know whether firewalls have formed by the Page-time, evaporation is all-important; by definition the Page time is has to do with evaporation. The remaining subsystem (a black hole in the case of interest) by the Page time will be described by thermal density matrix, and will have generic behavior. In particular, if a black hole has a firewall after the Page-time, then firewalls are generic features of black holes, i.e., they exist for the vast majority of black hole states.

On the other hand if we want to know whether a firewall has formed by the scrambling time, evaporation does not seem relevant. For an evaporating  black hole of mass $M,$ at the scrambling time the number of emitted quanta is only  $\log M,$  a negligible fraction of the total entropy. It is not evaporation, but rather the unitary evolution of a large system which causes scrambling.  Therefore we might as well ignore evaporation and assume the black hole is in a reflecting box.

The question then, is in what sense is has the pure state of a system become generic by the time it is scrambled? And does that degree of genericity imply the existence of a firewall?s



\setcounter{equation}{0}
\section{Fine Grained and Coarse Grained Operators}
\subsection{The Meaning of Fine and Coarse Grained}
Ordinarily, when dealing with a large system of many degrees of freedom, we are interested in coarse grained quantities. To illustrate the difference between  coarse and fine grained, consider a large system such as a box with perfectly reflecting walls. The box is filled with radiation and also some electrons to scatter the radiation and bring it to equilibrium. There are two cases to compare. In the first case the photons and electrons are put into the box in a pure quantum state with a given expectation value of the total energy. The quantum state at time zero is
\be
\Psi(0) = \sum_n F_n |n\rangle
\label{21}
\ee
where the index $n$ represents the $n^{th}$ energy eigenstate in the box.

 For convenience we will define the states $|n\rangle$ so that the $F_n$ are real. At a later time the state evolves to
\bea
\Psi(t) \eq \sum_n F_n e^{i\phi_n}  |n\rangle \cr
\eq  \sum_n F_n e^{iE_n t}  |n\rangle
\label{22}
\eea
The probability for the energy $E_n$ is $$P_n = F^{\ast}_n F_n.$$

The other situation assumes that the degrees of freedom in the interior of the box are entangled with a heat bath on the outside. We can imagine that the entanglement took place at a time when there was a hole in the box, which was subsequently sealed. We may assume that the density matrix has the form,
\be
\rho = \sum_n P_n |n\rangle \langle n |
\label{23}
\ee
at all times.

By fine-grained I mean that an observable is very sensitive to the relative phases between neighboring energy states as in \ref{22}. Coarse grained  means the opposite: a coarse grained quantity is insensitive ( exponentially small sensitivity in the size of the system) to those phases.  Coarse grained operators generally have practically the same expectation values in a pure state and in the mixed state \ref{23}.

For large closed systems  quantities built out of a small fraction of the degrees of freedom, will take on their coarse grained thermal values after a suitable scrambling time \cite{Hayden:2007cs} \cite{Sekino:2008he}. For example, take a sub-volume of a box filled with radiation, consisting of a small fraction of the total volume. To exponential precision all expectation values involving fields within the sub-volume tend to  the same value in  the pure and the mixed states. In fact it is thought that as long as the sub-volume is smaller than half the size of the box, that this remains so \cite{Page:1993wv}. Whenever this is true  the state is said to be scrambled  \cite{Hayden:2007cs} \cite{Sekino:2008he}.

On the other hand there are some quantities built out of more than half the system which are sensitive to the relative phases. Those that are, by definition, are fine-grained.

Obviously any quantity which can probe the purity of $|\Psi\rangle$ is fine-grained. Such quantities are  extremely complicated functions of at least half the degrees of freedom in the box.

The question arises: In  the description of a black hole in terms of microstates, is the existence of a firewall a fine-grained or coarse-grained issue? The answer that I propose is that it is very fine-grained.

\subsection{Special States and Generic States}

 Let's assume that the initial state has some special property. An example would be a reflecting box filled with highly coherent laser radiation. It is obvious that such a state is far from generic, and that the phases $\phi_n$ are special. It is also far from being scrambled. But as we will see,  scrambled and generic are different concepts: in particular scrambled does not imply generic.

 To see why they are different, first note that the scrambling time is typically polynomial in the number of degrees of freedom of the system. In the case of a black hole it is believed to be even shorter, namely logarithmic.

 Now consider the evolution of the phases in \ref{22}. In the initial special state the phases may be defined to be real. As time evolves they spread out on the unit circle. If the energies levels are characteristic of a chaotic system they will be incommensurate and will eventually be randomly sprinkled over the unit circle. In other words, the typical state will be characterized by a classical gas of distinguishable particles on the unit circle, with random unpredictable positions. The timescale for this to happen can be estimated by asking how long it takes for two neighboring phases $\phi_n$ and $\phi_{n+1}$ to separate by an order $1$ angle.

If we suppose that the entropy defined by the probability distribution $F^{\ast}_n  F_n $ is $S$ then the spacing between neighboring levels is of order
\be
\delta E \sim e^{-S}.
\label{2.4}
\ee

 After elapsed time $t$ the phase difference between neighboring energy levels will be
 \be
 \delta \phi \sim t \delta E = t e^{-S}
 \label{2.5}
 \ee

 The time scale for the phases to randomize\footnote{The randomizing of phases is not the same as de-coherence due to entanglement with an environment. However, they tend to have the same effect.}  will be the classical recurrence time,
 \be
 t_{rec} \sim e^S.
 \label{2.6}
 \ee

 By contrast, the scrambling time $t_{\ast}$ for a black hole of mass $M$ is only
 \be
 t_{\ast} = M \log M = \sqrt{S} \ \log{ S}
  \label{2.7}
 \ee
 At the scrambling time neighboring phases have only separated by an exponentially small amount,
 \be
  \delta \phi \sim t_{\ast} \delta E =  \sqrt{S} \ \log{ S} e^{-S}
   \label{2.8}
 \ee

Thus at the scrambling time the phases are  extremely coherent. Evidently  the scrambling time has nothing to do with the time for the state of a complex system to become generic.

 What kind of operators are sensitive to the relative  phases? By definition, not coarse grained operators. Here is an example of a fine-grained operator.
 \be
  {\cal{F} }= \sum_n   \left\{ \  |n\ra \la n+1 | + |n+1\ra \la n | \ \right\}
  \label{2.9}
 \ee

 Consider how the expectation value of $ {\cal{F} }$ varies with time.
 \be
 \langle \Psi(t)| {\cal{F} } | \Psi(t) \rangle = \sum_n F^{\ast}_n  F_{n+1}
 e^{i(E_n-E_{n+1}) t}  \ + \ cc
   \label{2.10}
 \ee
 where $E_n-E_{n-1}$ is of order $e^{-S}.$ For $t<< e^S$ the phase-factors can be ignored; they are extremely close to $1$. If we also assume that $F$ is a smooth function of $n$ then we see that $$ \langle \Psi(t)| {\cal{F} } | \Psi(t) \rangle \approx 1.$$

 However, as $t$ increases past the recurrence time, the phase differences between neighboring energies become random and
 $$ \langle \Psi(t)| {\cal{F} } | \Psi(t) \rangle \approx 0.$$

 This is the same value that the expectation value of ${\cal{O} }$ would have in the incoherent density matrix \ref{23}.

Note that nothing special happens at the scrambling time. At $t_{\ast}$ the neighboring phase differences are exponentially small and the expectation value of ${\cal{O} }$ is close to its value at $t=0.$

I have used the terms fine-grained and coarse-grained as if there is no middle ground, but there are varying degrees of fine-grained. The operator in \ref{2.9} is maximally fine grained because it depends on the phases of nearest-neighbor energy levels. If instead, the operator coupled second nearest neighbors the time scale for it to relax to zero would be more rapid.

 There are of course many other  highly fine-grained operators but \ref{2.8} is typical of them. In general they will achieve the value that they have in the incoherent density matrix only when the phases become random.

 By contrast, coarse grained observables tend to their incoherent counterparts much more rapidly; by the scrambling time.

\subsection{  Young Black Holes are Special}

\begin{figure}[h!]
\begin{center}
\includegraphics[scale=.3]{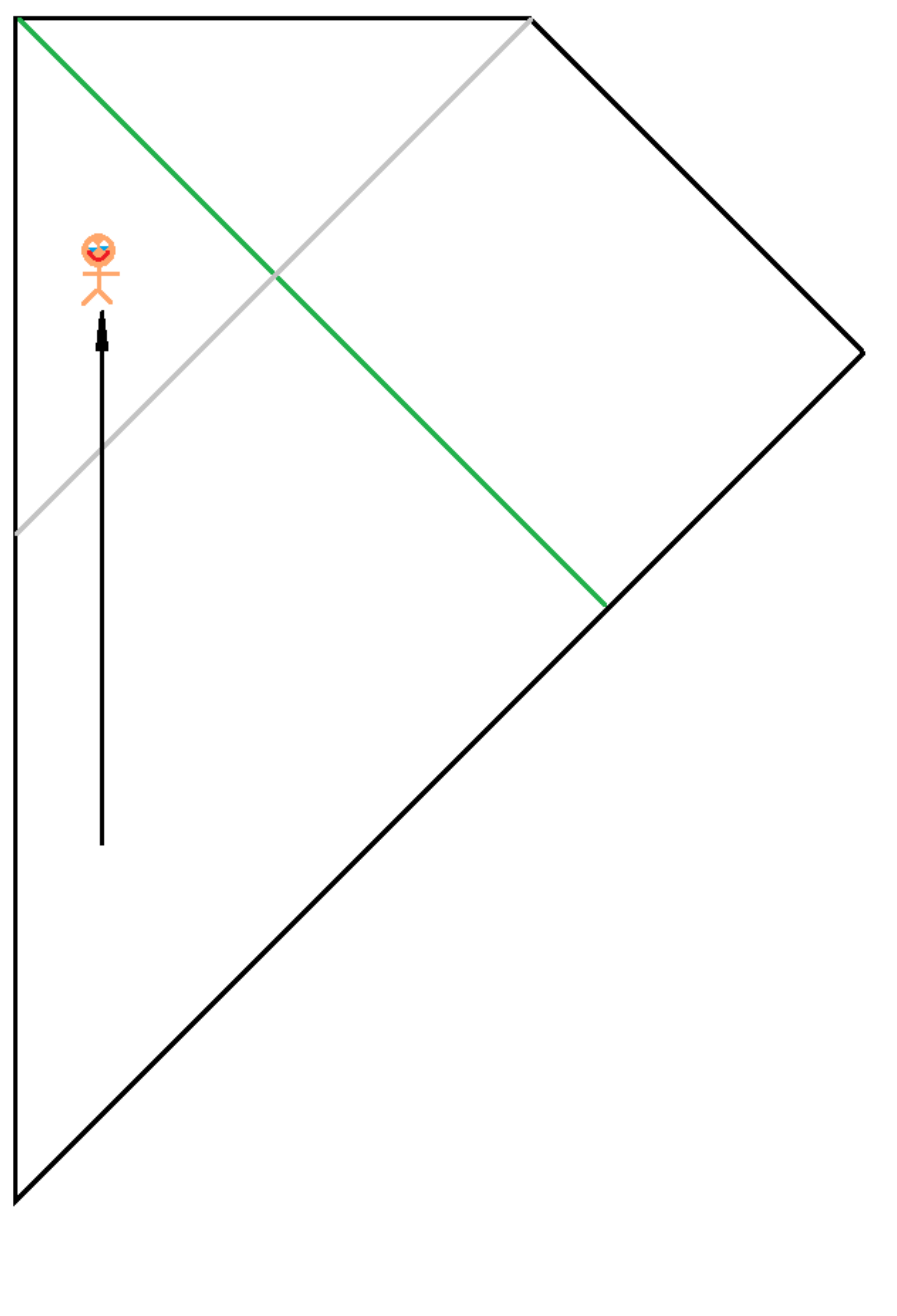}
\caption{Sufficiently young black holes must be safe. }
\label{f8}
\end{center}
\end{figure}

If it is true that the generic black hole state has a firewall, are there any special states that do not?  It is easy to argue that young black holes, formed in collapse, don't have firewalls. Consider the formation of a black hole by an in-falling shell as in Figure \ref{f8}.

 The true event horizon forms before the shell reaches the origin or even the Schwarzschild radius.
The figure shows the world line of an observer who enters the horizon while still out of causal contact with the shell. Locality insures that nothing happens when the observer crosses the horizon.

Does the observer survive passing through the in-falling shell? That is a matter of detail. In principle for a very large black hole the energy density of the shell can be arbitrarily small, and remain so  for a long time after it passes the trapped surface. Under such circumstances the observer will survive into the region beyond the shell, until tidal forces near the singularity become too large. Thus, a ``young" black hole shortly after formation cannot have a firewall.

We can now formulate a paradox. Consider the description of the black hole in the frame of an observer who remains outside the horizon.
Let's suppose that there exists a firewall-operator in the Hilbert space of black holes that detects the existence of a firewall. Call the firewall-operator ${\cal{F}}$ and define   it so that  the existence of a firewall is indicated by ${\cal{F}}=0$.  The arguments of \cite{Almheiri:2012rt} and \cite{Czech:2012bh}  imply that in any eigenstate of the Hamiltonian, a firewall exists. Therefore in the space of black hole states, ${\cal{F}}$ must  be identically zero.

But if this is so, then young black holes must have firewalls. But they don't. Evidently there are superpositions of energy eigenstates for which ${\cal{F} } \neq 0$ even though it is zero for every eigenstate.

The resolution of this paradox is that the Firewall operator is similar to the fine-grained operator (also called ${\cal{F} }$ ) in \ref{2.9}. The expectation values of that operator in almost all states (states with random phases) are very close to zero.

But in special states with smooth phase relations between neighbors, the expectation value of ${\cal{F}}$ is $\sim 1.$ Moreover, $\langle {\cal{F}} \rangle$ is time-dependent in the same way that the Firewall operator is. $\langle {\cal{F}} \rangle = 1$ for young black holes, and $\langle {\cal{F}} \rangle = 0$ for old black holes.

The question of how long it takes to form a firewall depends on just how fine-grained the operator ${\cal{F}}$ is. As discussed at the end of Section 2.2, the distinction between fine and coarse grained is not a sharp one. If ${\cal{F}}$ is maximally fine-grained (meaning that its matrix elements only couple nearest-neighboring energy levels) then it takes a very long time to form a firewall. In the case without evaporation that time is of order the recurrence time: in the case with radiation, the Page time.

One final point worth mentioning in regard to the special-ness of young black holes, is that the number of ways of making a black hole by collapse is probably much smaller that the exponential of the Bekenstein entropy. It has been pointed out by 't Hooft, and emphasized by Banks, that the entropy of ordinary matter which could collapse to form a black hole is much less than the black hole entropy. That fact supports the idea that young black holes are special states in the Hilbert space of collapsed objects.


\setcounter{equation}{0}
\section{What can Alice See? And What Can She Tell Bob?}
Although firewalls and migrating singularities seem to be a consistent hypothesis, some of the consequences are very strange and deserve careful thought before they are accepted. In this section I will explain my own reservations.

Consider Alice as she falls through the zone of an extremely large black hole.  We will follow her  from the time that she is at distance $D$ from the horizon until she crosses it, or is terminated at a firewall.              There is a tranverse area of order $D^2$ that contains everything that Alice can be in contact with before she passes the horizon. Because that area is small compared to the whole black hole, we can assume that it is described by a thermal density matrix. For young black holes we may either assume that the patch is maximally entangled with degrees of freedom behind the horizon, or, with the rest of the stretched horizon outside the patch. For an old black hole the entanglement is with the radiation $R.$ But in either case the density matrix of Alice's patch is exactly the same: thermal with temperature $1/{2 \pi}.$  Therefore there is no way by examining the degrees of freedom in the zone that she can tell if it is an old or young black hole.
What can she detect as she falls? There are fluctuations of the quantum fields  in the zone. In principle if she is equipped with suitable detectors she can detect these fluctuations.  If she is at distance $D$ she has time $D$ to make a measurement. The simplest thing she can measure is the value of some  field smeared over a volume $D^3.$ If the field is weakly coupled, its probability distribution  will be approximately Gaussian  (assume it is centered at zero).  Moreover the probability  is exactly the same for young and old black holes. Since we do not expect Alice to experience a sense of burning or a flux of high energy particles in the zone of a young black hole, she will not have these experiences in the old black hole, at least before she reaches the horizon.

Now suppose the black hole is old, and Alice has had contact with the Hawking radiation before entering the zone. She has measured an observable in the radiation that is entangled with a field fluctuation in a particular volume $D^3$ at distance $D.$ As she passes the volume she measures $D.$ If the zone is really entangled with $R$ she will see the field have the value she expects. Such measurements may be difficult and not completely precise, but  the probability distribution will be biased away from zero.

Alice  cannot experience any fluctuation which she could not have experienced without measuring $R,$ but now she knows in advance what she will experience. Let's consider something highly improbable such as Alice getting hit by a high energy photon. The probability of that happening is no higher than if Alice did not look at the radiation. The only new thing is that if, with low probability, there is a very high energy photon in her path, she will know it before she reaches it.

By gathering statistics of correlations between $R$ and the fields in the zone, Alice will know with high certainty whether the black hole is old enough to have a firewall. But if she does not do such correlated measurements, the only indication of the firewall is at the point where she reaches the horizon. If the black hole is young she will find herself on the other side unharmed. If the black hole is old there is no other side, and Alice is terminated instantly,  without any warning, and without being able to confirm the existence of the firewall. The knowledge of the firewall seems to have no predictive implications. This prediction of a non-predictive firewall is extremely odd.

Bousso \cite{Bousso:2012as} and Harlow \cite{Harlow:2012me} have suggested a very interesting resolution of the firewall puzzle, which says that each observer has their own theory, appropriate to their own causal patch, and that the theories or observations do not have to agree as long as different observers cannot communicate and discover a contradiction.  The idea is called \it strong complementarity. \rm
In particular Harlow and Bousso argue that the theory in Alice's causal patch will not predict correlation between $R$ and the fluctuation in the zone. Bob's theory, by contrast, does predict such correlations (remember that both $R$ and zone are visible to Bob). That sounds far-fetched, but we may ask whether it leads to a contradiction. This would be the case if Alice could communicate to Bob that she did not see the fluctuation which Bob's theory predicts. Thus the Bousso-Harlow theory revolves around the possibility that Alice can measure a field fluctuation, and still have time to report the result to Bob. Bob she can then strong complementarity is falsified.

The experimental question  is very subtle. Alice must do it without disturbing the system enough to invalidate the result. There are many requirements for such an experiment and most of them create the danger that a soft graviton will be emitted and fall in to the stretched horizon. If it does so a scrambling time before Alice makes her measurement, it will disturb the entanglements that Alice is attempting to measure.

If it is possible to prove that no experiment can be done which allows Alice to contradict Bob about entanglement, then the Bousso Harlow theory of strong complementarity may be a viable alternative to firewalls.

\setcounter{equation}{0}
\section{Conclusion}

The various arguments for firewalls are not completely conclusive but they do seem plausible.
The different pieces have varying degrees of plausibility. In the case of old black holes, the low angular momentum  argument  uses the entanglement between modes of the asymptotic Hawking radiation, and the use of low energy field theory outside the stretched horizon. It does not depend on the existence of a microscopic quantum description of the black hole, but it does depend on being able to trace the outgoing modes back into the zone behind the centrifugal barrier. It has the highest degree of plausibility.

For the high angular momenta that make up most  of the entropy of the horizon, more than just the S-matrix is needed. There may be loopholes; for example it may not be possible to separate the degrees of freedom of the stretched horizon from  the field degrees of freedom between the horizon and the centrifugal barrier.  If one does accept the maximal entanglement argument, then the horizon, after the page time, becomes a firewall. In that case complementarity does not apply to it. The most consistent view may be that once a firewall forms, it should no longer be called a horizon.
With that view firewalls and complementarity are both correct. The only question would be whether black holes ever do have horizons? As we've seen, the answer is yes; sufficiently young black holes  must have horizons.

The firewall is not observable to Alice as she falls toward the horizon. What is observable, if she cares to measure them, is correlations between the Hawking radiation $R$ and field fluctuations in the zone. If she detects such fluctuations she may realize that the horizon is dangerous and try to avoid it. But if she makes no intricate measurements on $R$ outside the black hole, there is no way for Alice to know that there is a firewall in her future. This leads to the odd fact that the firewall has no predictive implications.

On the other hand, if she falls in somewhat before the Page time, she may detect that the geometry is non-standard after she passes the horizon and before she hits the singularity. 

An important question raised in \cite{Almheiri:2012rt} is how long the horizon of black hole lasts? The authors claim that what the maximal entanglement shows is that the typical black hole state has a firewall. This is certainly correct. But they go on to say that the state of a black hole becomes typical after the scrambling time.  That would suggest that the Page time and Hawking evaporation  ware merely crutches, and that firewalls would occur after the scrambling time, even without evaporation.

The counter argument to this last claim is that the scrambling time is not the time for
the state of a system to become typical. It is the time for subsystems smaller than half the system to become typical. There are many subtle global observables that do not become typical until much longer times. If the existence of a firewall is one of these subtle questions then the timescale for the formation can be long.

It should be pointed out that both the Page and scrambling times very long in a certain sense. They both go to infinity as $\hbar$ goes to zero, although the scrambling time  only logarithmically.

One  point that was emphasized is that firewalls are about apparent horizons. They only merge with the true horizon after the last bit of matter has fallen in. An observer can always enter the true horizon, but the survival time becomes extremely short after the Page time.

Finally, there is one loophole that is especially interesting. Bousso and Harlow have argued for a strong form of complementarity in which different observers would have apparently contradictory theories of entanglement as long as they cannot communicate a contradiction. The viability of strong complementarity would depend on a no-go theorem for an experiment in which Alice determines the absence of entanglement, and communicates this to Bob. At the moment there is no such no-go theorem.

\section*{Acknowledgements}

I thank Don Marolf for an early explanation of the argument of \cite{Almheiri:2012rt}, and
Joe Polchinski for  a preliminary version of the paper well before publication.

I am also grateful to Raphael Bousso Daniel Harlow, Steve Shenker, and Douglas Stanford for many discussions and insights into the firewall phenomenon. That does not mean that they necessarily agree with my views.

\end{document}